\documentclass[%
 reprint,
superscriptaddress,nofootinbib,
 amsmath,amssymb,
 aps,
]{revtex4-2}

\usepackage{graphicx}% Include figure files
\usepackage{dcolumn}% Align table columns on decimal point
\usepackage{bm}% bold math
\usepackage{enumitem}
\usepackage{xcolor}

\usepackage{amsmath}
\DeclareMathOperator*{\argmax}{argmax}
\begin{document}

\preprint{APS/123-QED}

\title{Learn your entropy from informative data:\\ an axiom ensuring the consistent identification of generalized entropies}% Force line breaks with \\

\author{Andrea Somazzi}
\email{andrea.somazzi@sns.it}
 \affiliation{IMT School for Advanced Studies, Piazza S. Francesco 19, 55100 Lucca, Italy}
 \affiliation{Scuola Normale Superiore, Piazza dei Cavalieri 7, 56126 Pisa, Italy}%Lines break automatically or can be forced with \\
\author{Diego Garlaschelli}
\affiliation{IMT School for Advanced Studies, Piazza S. Francesco 19, 55100 Lucca, Italy}
\affiliation{Lorentz Institute for Theoretical Physics, Niels Bohrweg 2, 2333 CA Leiden, The Netherlands}
\affiliation{INdAM-GNAMPA Istituto Nazionale di Alta Matematica, Italy}


\begin{abstract}
Shannon entropy, a cornerstone of information theory, statistical physics and inference methods, is uniquely identified by the Shannon-Khinchin or Shore-Johnson axioms.
Generalizations of Shannon entropy,  motivated by the study of non-extensive or non-ergodic systems, relax some of these axioms and lead to entropy families indexed by certain `entropic' parameters. 
In general, the selection of these parameters requires pre-knowledge of the system or encounters inconsistencies. Here we introduce a simple axiom for any entropy family: namely, that no entropic parameter can be inferred from a completely uninformative (uniform) probability distribution. 
When applied to the Uffink-Jizba-Korbel and Hanel-Thurner entropies, the axiom selects only R\'enyi entropy as viable. It also extends consistency with the Maximum Likelihood principle, which can then be generalized to estimate the entropic parameter purely from data, as we confirm numerically.
Remarkably, in a generalized maximum-entropy framework the axiom implies that the maximized log-likelihood always equals minus Shannon entropy, even if the inferred probability distribution maximizes a generalized entropy and not Shannon's, solving a series of problems encountered in previous approaches.
\end{abstract}

\maketitle

\section{Introduction} 
The concept of entropy was introduced by Clausius in the thermodynamic framework \cite{clausius1856xxxiii} and later adopted in statistical physics by Boltzmann and Gibbs as a tool to describe macroscopic systems in terms of their probabilities of occupancy of microscopic states \cite{boltzmann1877beziehung, gibbs2014elementary}. Within information theory, Shannon axiomatically (re)derived the entropy as a quantification of the uncertainty encoded in a probability distribution, applicable to (among other things) the compressibility of sequences of symbols generated by ergodic probabilistic sources~\cite{shannon1948mathematical}.
This allowed Jaynes to subsequently propose that the distribution that maximizes Shannon entropy, under the constraints implied by the empirical information available about a real system, provides the least biased (maximally noncommittal) inferential description of the unknown microscopic details of that system, a construction that can be used to reinterpret  statistical physics from an information-theoretic viewpoint~\cite{jaynes1957information}.
In modern research, statistical inference and model identification based on entropy maximization are perfectly consistent with Maximum-Likelihood estimation methods and are at the heart of several machine-learning techniques~\cite{machinelearning}.\\

Various generalizations of Shannon entropy (the most popular of which were motivated by the statistical physics of non-extensive and/or non-ergodic systems) have been proposed in various contexts~\cite{tsallis2009introduction,thurner,amigo2018brief, lopes2020review}, resulting in  extended families of entropy that depend, besides the usual parameters, on extra `entropic' parameters. 
For a fixed choice of these parameters, one can still maximize the resulting entropy and generalize the inference procedure.
This is possible when there is enough knowledge \emph{a priori} about the system, so that the entropic parameter can be set `by hand' to the correct value.
However, it is generally not possible to maintain compatibility with the Maximum-Likelihood principle and, crucially, to infer the values of the entropic parameters purely from data without encountering inconsistencies, making the generalized methodology inapplicable without prior knowledge of the correct entropy.\\

In this paper we discuss and alleviate those inconsistencies by introducing an axiom that restricts the form of parametric entropies. The axiom enforces a simple information-theoretic requirement and allows for the consistent inference of the entropic parameters purely from the available data, as we show via analytical results and numerical examples.
The paper is organized as follows. In Sec.~\ref{background} we first review the theoretical background behind the axiomatic definitions of entropy, the Maximum Entropy Principle, and the Maximum Likelihood Principle.
In Sec.~\ref{results} we then discuss the main contributions of the paper, i.e. the introduction of the new axiom and its implications for the selection of entropies from certain popular families, the restoration of consistency with the Maximum-Likelihood principle, and the generalization of the latter in order to infer the entropic parameter(s) purely from the data. 
Finally, in Sec.~\ref{conclusions} we offer some concluding remarks.

\section{Theoretical background\label{background}}
\subsection{The Shannon-Khinchin axioms}
Given a  distribution $P=(p(G_1),\dots,p(G_\Omega))$, where $p(G_i)$ is the probability that the random variable $G$ takes the $i$-th discrete outcome (or `state') $G_i$, $\Omega$ is the total number of distinct outcomes, and clearly $\sum_{i=1}^\Omega p(G_i)=1$, Shannon entropy $S[P]$ is axiomatically defined through the following four Shannon-Khinchin (\emph{SK}) axioms \cite{khinchin1957mathematical}:

\begin{itemize}
\item \textit{SK1 (continuity):} $S[P]$ is continuous in the entries of $P$. 
\item \textit{SK2 (maximality):} $S[P]$ is maximal when $P$ is the uniform distribution $P_u\equiv({\Omega}^{-1},\dots,{\Omega}^{-1})$. 
\item \textit{SK3 (expansibility):} $S[P]$ is expansible, i.e. it does not change if for the variable $G$ an $(\Omega+1)$-th outcome with zero probability ($p(G_{\Omega+1})=0$) is added: $$S[(p(G_1),\dots,p(G_\Omega))]=S[(p(G_1),\dots,p(G_\Omega),0)].$$ 
\item \textit{SK4 (separability):} the entropy of the joint distribution $R=(r(G_1,G'_1),\dots, r(G_\Omega,G'_{\Omega'}))$ of two variables $G$ and $G'$ with marginal distributions $P=(p(G_1),\dots,p(G_\Omega))$ and $Q=(q(G'_1),\dots,q(G'_{\Omega'}))$ respectively, where $p(G_i)=\sum_{j=1}^{\Omega'} r(G_i,G'_j)$ and $q(G'_j)=\sum_{i=1}^\Omega r(G_i,G'_j)$, separates as
$$S[R]=S[P]+S[Q|P].$$ 
Here $S[Q|P]$ is the conditional entropy of $Q$ on $P$, defined as $S[Q|P]=\sum_{k=1}^\Omega p(G_k) S[Q_{|k}]$ with  $Q_{|k}=({r(G_k,G'_1)}/{p(G_k)},\dots,{r(G_k,G'_{\Omega'}}/{p(G_k)})$ denoting the conditional distribution of the events in $Q$ on the $k$-th event in $P$. 
Note that in particular, if the two events are independent ($Q=Q_{|k}$ for all $k$), then $S[R]=S[P]+S[Q]$, in which case separability becomes \emph{additivity}.\\
\end{itemize}

It is possible to show that, up to an inessential overall multiplicative factor, the only functional form of $S[P]$ respecting the four \emph{SK} axioms is Shannon entropy:
\begin{equation}
    S_1[P]=-\sum_{i=1}^\Omega p(G_i)\ln p(G_i),
    \label{S1}
\end{equation}
where the subscript $1$ will be justified later. 
As required by \emph{SK2}, the maximum value of $S_1[P]$ is attained by the uniform distribution $P_u$, leading to Boltzmann entropy:
\begin{equation}
S_1[P_u]=\ln\Omega.
\label{boltzmann}
\end{equation}
No distribution $P$ can be such that $S_1[P]>S_1[P_u]$.

\subsection{The Maximum Entropy Principle\label{MEP}} 
The informational entropy $S_1[P]$ in Eq.~\eqref{S1} coincides with the physical entropy derived by Gibbs~\cite{ gibbs2014elementary}, which in turn generalizes Boltzmann entropy in Eq.~\eqref{boltzmann}~\cite{boltzmann1877beziehung}.
This equivalence is not coincidental and is rooted in statistical inference, as Jaynes showed with the introduction of the Maximum Entropy Principle (MEP) \cite{jaynes1957information}. 
The MEP states that, given only a set $I$ of pieces of empirical information about a system (in the physical situation, this typically means the knowledge of a few, macroscopic conserved quantities such as the total energy and/or the total number of particles), one should assign the possible microscopic states a probability distribution $P$ that maximizes the entropy. 
This maximum-entropy distribution is sometimes denoted as $P=\circ I$. 
In other words, entropy can be used as an inference functional whose maximization minimizes bias and prevents arbitrariness.\\

In particular, consider a system with a set of $\Omega$ potential microstates $\{G_i\}_{i=1}^\Omega$ and assume that the available information $I$ is encoded in the empirical value $C^*=C(G^*)$ of a certain (scalar or vector) function $C$ of the microstate of the system, where $G^*$ is the particular (unobservable) empirical microstate. $C^*$ is the only observation available.
Since $G^*$ is unknown, the microstate is treated as a random variable $G$.
The MEP applied to $S_1[P]$ identifies the maximum-entropy distribution for $G$, which we denote as $P_0=(p_0(G_1),\dots,p_0(G_\Omega))$ or $P_1=(p_1(G_1),\dots,p_1(G_\Omega))$, depending on whether $C^*$ is treated as a `hard' or `soft' constraint, respectively.\\

In the case of hard constraints (\emph{microcanonical ensemble}), only a restricted number $\Omega_{C^*}<\Omega$ of microstates $i$ for which $C(G_i)$ matches $C^*$ exactly are assigned a non-zero probability, which (due to \emph{SK2} and \emph{SK3}) has to be uniform over the restricted support, i.e. $p_0(G_i)=\Omega^{-1}_{C^*}$ if $C(G_i)=C^*$ and $p_0(G_i)=0$ otherwise.
The resulting entropy is \begin{equation}
S_1[P_0]=\ln\Omega_{C^*}< S_1[P_u].
\label{unif2}
\end{equation}
Unfortunately, calculating $\Omega_{C^*}$ is generally a hard combinatorial problem, which makes the microcanonical ensemble not amenable to analytical calculations.\\

In the case of soft constraints (\emph{canonical ensemble}), only the expected value $\langle C \rangle$ of the observable is constrained to match $C^*$, i.e.
\begin{equation}
    \langle C \rangle\equiv\sum_{i=1}^\Omega p(G_i)\,C(G_i)=C^*,
    \label{mean}
\end{equation}
thus allowing for the full set of $\Omega$ microstates, however with a non-uniform probability $p(G_i)$ yet to be determined.
To find the specific probability $p_1(G_i)$ maximizing $S_1$ under the soft constraint above, one can introduce the Lagrange multiplier $\theta$ (which has the same dimensionality as $C$), plus an additional multiplier $\alpha$ enforcing the normalization of $P$, and look for the specific values (denoted as $P_1,\theta_1,\alpha_1$) for which all the derivatives of the Lagrangian function
\begin{equation}
\label{mepq}
\mathcal{L}_1[P]\equiv S_1[P]-\alpha\left[\sum_{i=1}^\Omega p(G_i)-1\right]-\theta\cdot\left[\langle C \rangle - C^*\right]
\end{equation}
vanish (the notation $\theta\cdot C$ indicates the scalar product).
Setting $\partial \mathcal{L}[P]/\partial P|_{P_1}=0$, i.e. $\partial \mathcal{L}[P]/\partial p(G_i)|_{p_1(G_i)}=0$ $\forall i$, leads to the functional form of $P_1$, which turns out to be the well-known Boltzmann-Gibbs distribution with entries
\begin{equation}    p_1(G_i,\theta)=\frac{ e^{-\theta\cdot C(G_i)}}{Z_1(\theta)},\quad Z_1(\theta)=\sum_{j=1}^\Omega e^{-\theta\cdot C(G_j)},
\label{BG}
\end{equation} where $Z_1(\theta)$ is the \emph{partition function}, resulting from the normalization constraint 
\begin{equation}
\left.\frac{\partial \mathcal{L}[P_1]}{\partial \alpha}\right|_{\alpha_1}=0\quad\Rightarrow\quad\sum_{i=1}^\Omega p_1(G_i,\theta)\,=1
\label{alpha1}
\end{equation}
which leads to
\begin{equation}
    \alpha_1=-1+\ln Z_1(\theta)
    \label{alpha1bis}
\end{equation}
independently of the value of $\theta$.\\

Importantly, $P_1$ is not identified entirely, until the parameter $\theta$ is also determined. This is attained by enforcing the vanishing of the remaining derivatives, identifying the value $\theta_1$ realizing Eq.~\eqref{mean}:
\begin{equation}
\left.\frac{\partial \mathcal{L}[P_1]}{\partial \theta}\right|_{\theta_1}=0\quad\Rightarrow\quad\sum_{i=1}^\Omega p_1(G_i,\theta_1)\,C(G_i)=C^*,
\label{theta1}
\end{equation}
where, if $\theta$ is a vector, the notation means again that all the derivatives of $\mathcal{L}[P]$ with respect to the components of $\theta$ vanish separately.
The final solution to the MEP problem is therefore given by inserting $\theta_1$ into Eq.~\eqref{BG}, and we will denote it as $P_1(\theta_1)=(p_1(G_1,\theta_1),\dots,p_1(G_\Omega,\theta_1))$.
The MEP with soft constraints, which are appropriate when the observables are expected to fluctuate, has been  used successfully for inference and model selection in many fields beyond physics, including network theory, neuroscience, economics and biology \cite{karmeshu2003entropy,squartini2017maximum}.

\subsection{The Maximum-Likelihood Principle\label{sec:ML}}
It is very important to realize that the MEP procedure outlined above has deep connections and desirable consistencies with the Maximum Likelihood (ML) principle, which applies to more general (not necessarily maximum-entropy) parametric probability distributions and states that the optimal parameter value $\theta^*$ is the one maximizing the log-likelihood on the data $G^*$.
In our setting, the ML principle would look at Eq.~\eqref{BG} as any other parametric distribution and select the value 
\begin{equation}
\theta_1^*=\argmax_\theta {\ell_1(\theta)},\quad %\textrm{where}\quad 
\ell_1(\theta)\equiv\ln p_1(G^*,\theta).
\label{ML}
\end{equation}
As a first result, it is easy to show that the value $\theta_1^*$ defined by Eq.~\eqref{ML} coincides with the value $\theta_1$ defined by Eq.~\eqref{theta1}~\cite{likelihood,garlaschelli2008maximum}, i.e.  $\theta_1^*\equiv\theta_1$ (in our notation, the asterisk next to a parameter will always denote the ML value of that parameter), i.e.
\begin{equation}
\left.\frac{\partial \ell_1(\theta)}{\partial \theta}\right|_{\theta^*_1}=0\quad\Rightarrow\quad\sum_{i=1}^\Omega p_1(G_i,\theta^*_1)\,C(G_i)=C^*
\label{theta1*}
\end{equation}
in analogy with Eq.~\eqref{theta1}.
This means that the ML principle can be seen as equivalent to the part of the Lagrangian optimization relative to $\theta$.\\

Moreover, it is straightforward to show that the maximized log-likelihood equals minus the entropy:
\begin{equation}
S_1[P_1({\theta^*_1})]=-\ell_1(\theta^*_1),
\label{Shannonlikelihood}
\end{equation} 
which is the counterpart of Eq.~\eqref{unif2} in the case of soft constraints.
This relationship is very important, because the maximized likelihood is at the basis of model selection criteria~\cite{Burnham,MDL}: if alternative models (i.e. alternative parametric probability distributions) are compared against the same empirical data, the model to be preferred (assuming all models have the same complexity, e.g. the same number of parameters) is the one with highest maximized likelihood.
Then, Eq.~\eqref{Shannonlikelihood} ensures that the ranking of models based on ML is the same as the ranking based on minus their entropy: the least uncertain (i.e. most informative) model has to be preferred.
For models with different numbers of parameters and/or functional forms, the ranking based on likelihood/entropy has to be revised   by adding a term controlling for the variable model complexity, leading to criteria such as AIC, BIC, the Minimum Description Length, etc.~\cite{Burnham,MDL} (for simplicity, we will not consider this situation here).
Also note that, when the maximum-entropy distribution $P_1(\theta_1^*)$ is inserted into Eq.~\eqref{mepq}, we get
\begin{equation}
\mathcal{L}_1[P_1(\theta_1^*)]=S_1[P_1(\theta_1^*)]=-\ell_1(\theta_1^*),
\label{lagranlike}
\end{equation}
so that the Lagrangian, evaluated at $P_1(\theta_1^*)$, coincides with minus the maximized log-likelihood, and can therefore be used to rank alternative models as well. 
All the above results indicate that the MEP can be used as a model selection criterion, exactly as the ML principle, by ranking models based on their realized entropy.\\

It is also important to consider the case when there are $M$ independent observations $\{C^*_m\}_{m=1}^M$ about the system, which technically means that there are $M$ independent and identically distributed (i.i.d.) realizations $\{G^*_m\}_{m=1}^M$ of the microstate $G$ (recall that $G$ is treated as a random variable), on each of which the quantity $C^*_m=C_m(G^*_m)$ ($m=1,M$) is observed. 
Clearly, since the system being observed multiple times is one, the probability distribution characterizing it must still be specified by a single value of the Lagrange multiplier $\theta$ coupled to the quantity $C$.
It should at this point be noted that the principle that identifies how to optimally combine the $M$ observations $\{C^*_m\}_{m=1}^M$ in order to estimate $\theta$ is not the MEP, but the ML one. Indeed, the ML principle applied to the joint log-likelihood $\sum_{m=1}^M \ln p_1(G^*_m,\theta)$, or equivalently to the average log-likelihood $\overline{\ell}_1(\theta)\equiv\sum_{m=1}^M \ln p_1(G^*_m,\theta)/M$, can be formulated by replacing Eq.~\eqref{ML} with
\begin{equation}
\theta_1^*=\argmax_\theta {\overline{\ell}_1(\theta)},\quad %\textrm{where}\quad 
\overline{\ell}_1(\theta)\equiv\frac{\sum_{m=1}^M\ln p_1(G_m^*,\theta)}{M}.
\label{MLM}
\end{equation}
It is easy to show that the condition $\partial \overline{\ell}_1(\theta) /\partial\theta|_{\theta^*_1}=0$ identifying $\theta_1^*$ leads to the well-known result
\begin{equation}
\langle C\rangle=\frac{1}{M}\sum_{m=1}^M C^*_m
\label{arithm}
\end{equation}
where the (arithmetic) sample average of the $M$ observations has emerged. 
So, in order to find the ML parameter value $\theta_1^*$, one should replace Eq.~\eqref{mean} with Eq.~\eqref{arithm}, or equivalently redefine $C^*$ in Eq.~\eqref{mean} as the sample average of $\{C^*_m\}_{m=1}^M$.
In plain words, the sample average is `produced' by the ML principle. 
On the contrary, within the MEP construction, there is no way of `telling' Eqs.~\eqref{mepq} and~\eqref{theta1} what, in case of $M$ observations, the meaning and definition of $C^*$ should be.
So in this case the ML principle is more informative than the MEP, and this is another reason why one wants the entropy to be fully consistent with what the ML principle leads to. 
In particular, it is easy to show that, due to the independence of the $M$ samples, the maximized average log-likelihood $\overline{\ell}_1(\theta_1^*)$ is still equal to minus the entropy:
\begin{equation}
S_1[P_1({\theta^*_1})]=-\overline{\ell}_1(\theta^*_1),
\label{ShannonlikelihoodM}
\end{equation} 
generalizing Eq.~\eqref{Shannonlikelihood}.
Note that there is no microcanonical counterpart of Eq.~\eqref{ShannonlikelihoodM}, since Eq.~\eqref{unif2} cannot be generalized to the case $M>1$, unless all the $M$ values $\{C^*_m\}_{m=1}^M$ are identical.
Indeed the microcanonical ensemble cannot be constructed, because by definition it cannot account for different realizations of the values of the constraints:
in case of different  observations of the same constraints, only the canonical ensemble is feasible.\\

The above discussion clarifies that it is important that the entropy is consistent with the maximized log-likelihood, because the ML principle is needed both for model selection and for the determination of how multiple observations of the same system should be combined in order to optimally estimate the parameters.

\subsection{The Shore-Johnson axioms}
An alternative axiomatic definition of an entropy functional, whose maximization in presence of a set $I$ of pieces of information should lead to a probability distribution $P=\circ I$ with certain properties, was proposed by Shore and Johnson (\emph{SJ}) through the following axioms \cite{shore1980axiomatic}:
\begin{itemize}
\item \textit{SJ1 (uniqueness):} given $I$,  $P=\circ I$ is unique. 
\item \textit{SJ2 (invariance):} if $\Gamma[\cdot]$ is a coordinate transformation (change of variables), then $\Gamma[\circ I]=\circ(\Gamma[I])$. 
\item \textit{SJ3 (system independence):} given two independent systems $A$ and $B$, it should not matter whether one accounts for distinct pieces of information about  them separately (in terms of marginal probabilities) or jointly (in terms of a joint probability). This means $\circ(I_A \wedge I_B)=(\circ I_A)(\circ I_B)$, where $I_A \wedge I_B$ denotes the union of the available pieces of information $I_A$ and $I_B$ about $A$ and $B$ respectively. 
\item \textit{SJ4 (subset independence):} it should  not matter whether one treats an independent subset of system states in terms of a separate conditional density or in terms of the full system density. 
Consider a partition of the system's states into disjoint subsets $\{\Lambda_k\}_k$ such that $\bigcup_k \Lambda_k = \Omega$, for each $k$ of which there is a piece of information $I_k$ available. Then $(\circ I)_{\Lambda_k}=\circ I_k$ $\forall k$, where $I=\bigwedge_k I_k$ is the total information, and $P_{\Lambda_k}=(p_{\Lambda_k}(G_1),\dots,p_{\Lambda_k}(G_\Omega))$, where $p_{\Lambda_k}(G_i)=p(G_i|G_i\in\Lambda_k)$ denotes the  conditional distribution relative to the subset $\Lambda_k$. 
\item \textit{SJ5 (maximality)\footnote{Actually, Shore and Johnson defined the maximality axiom only implicitly. Indeed, starting from the principle of minimum cross-entropy, they introduced the MEP as its equivalent in the case where the prior distribution is uniform. For this reason, even if not explicitly axiomatized, they considered the posterior $P$ to be equal to the uniform distribution (i.e. the same as the prior) when no information is available.}:} when no information is available ($I=\varnothing$), $P=\circ I$ is the uniform distribution $P_u$.\\
%\end{SJ}
\end{itemize}

Shore and Johnson claimed that Shannon entropy is the only inference functional compatible with their axioms, a statement suggesting the equivalence of the \emph{SK} and the \emph{SJ} axioms. However, it was later clarified~\cite{uffink1995can, jizba2019maximum} that Shore and Johnson's conclusion was due to an additional hidden assumption they made inadvertently when formally using \emph{SJ3} in their reasoning.   
Specifically, they considered a situation where distinct pieces of information $I_A$ and $I_B$ are known about two systems $A$ and $B$, and implied that the resulting joint probability factorises as $\circ(I_A \wedge I_B)=(\circ I_A)(\circ I_B)$, thereby applying \emph{SJ3} even if the independence of the two systems is not guaranteed (having only disjoint pieces of information about two systems does not guarantee that the two systems are independent)~\cite{uffink1995can, jizba2019maximum}. 
The presence of this additional assumption implies that Shannon entropy is in fact the desired functional only when systems are  independent: if this is not the case, then the resulting maximum entropy distribution is no longer `maximally non committal with respect to missing information', as Jaynes' MEP demands it to be~\cite{jaynes1957information}, because there is actually no `information' available about the (in)dependence of the systems.

\subsection{Generalized entropies}
Uffink~\cite{uffink1995can} showed that, if Shore and Johnson's proof is correctly revisited without the extra unjustified assumption, the entropy resulting from the \emph{SJ} axioms is not uniquely determined and is actually an entire generalized family $S^{(f)}_q[P]$, given by any increasing function $f$ of a certain functional $U_q[P]$ that we will call the Uffink functional, i.e. \begin{equation}
S^{(f)}_q[P]=f(U_q[P]),\quad U_q[P]=\bigg(\sum_{i=1}^\Omega p^q(G_i)\bigg)^\frac{1}{1-q}
\label{uffink}
\end{equation}
for some parameter $q>0$.
For a given $f$, each entropy in the family is identified by the parameter $q$, which we will therefore call the `entropic parameter'. Note that an entropic parameter plays a different role with respect to other structural parameters entering the entropy, such as $\theta$ in the Shannon case discussed above. 
Clearly, Shannon entropy must be one of the possible members of this family, and indeed, taking $f(x)= \ln x$, one can show that
\begin{eqnarray}
\lim_{q\to 1} S^{(\ln)}_q[P]&=&\lim_{q\to 1}\ln U_q[P]\nonumber\\ 
&=&-\sum_{i=1}^\Omega p(G_i) \ln p(G_i)\nonumber\\
&=&S_1[P].\label{wShannon}
\end{eqnarray}
In other words, Shannon entropy formally corresponds to $q=1$, justifying the subscript adopted in Eq.~\eqref{S1} (note instead that the subscript in the uniform distribution $P_0$ used in Sec.~\ref{MEP} to describe the microcanonical distribution under hard constraints has nothing to do with the case $q=0$, which is inadmissible).
Notably, Jizba and Korbel~\cite{jizba2019maximum,jizba2020shannon} showed that an entropy of the type $f(U_q[P])$ can also be obtained from the \emph{SK} axioms, provided that \emph{SK4} is relaxed to a generalized separability condition where the sum is replaced by the so-called Kolmogorov-Nagumo sum\footnote{Considering a bijection $f^{-1}:M\mapsto N\subset \mathbb{R}$, the generalized arithmetics is defined as follows:
\begin{eqnarray*}
    x \oplus y &=& f(f^{-1}(x)+f^{-1}(y)), \\
    x \ominus y &=& f(f^{-1}(x)-f^{-1}(y)), \\
    x \otimes y &=& f(f^{-1}(x)f^{-1}(y)), \\
    x \oslash y &=& f(f^{-1}(x)/f^{-1}(y)).
\end{eqnarray*}}, previously introduced in the context of generalized arithmetics~\cite{kolmogorov1930notion, nagumo1930klasse}. 
This shows that the \emph{SJ} axioms are actually equivalent to a specific generalization of the \emph{SK} ones. The generalization of \emph{SK4} has been a matter of discussion in the statistical physics literature for decades, as it relates to the subject of non-extensive (or rather non-additive) thermodynamics~\cite{tsallis2009introduction}.
We will call any entropy of the form $f(U_q[P])$ an Uffink-Jizba-Korbel (\emph{UJK}) entropy.\\

Several other generalized families of entropy resulting from relaxations of the \emph{SK} or \emph{SJ} axioms have been proposed~\cite{amigo2018brief, lopes2020review}.
A notable example is the so-called $(c,d)$-entropies $S_{c,d}[P]$ introduced by Hanel and Thurner \cite{hanel2011comprehensive} by replacing \emph{SK4} with the assumption of \emph{trace-form} (or more in \emph{general composable}) entropies, i.e. entropies that can be written as (functions of) a sum over the states $\{G_i\}_{i=1}^\Omega$ of the system. In particular, an entropy $S(P)$ is trace-form if it can be written as a sum $\sum_{i=1}^\Omega g\left(p(G_i)\right)$
for some function $g$. 
Note that Shannon entropy is in this class, with $g(x)=-x\ln x$.
More generally, a composable entropy can be written as a function $h$ of such a sum, i.e. 
\begin{equation}
S^{(h,g)}_{c,d}[P]=h\left(\sum_{i=1}^\Omega g\left(p(G_i)\right)\right),
\label{composable}
\end{equation}
where the entropic parameters ($c,d$) are determined by how the entropy scales with the number $\Omega$ of accessible configurations~\cite{thurner,hanel2011comprehensive}.
In particular, one considers the transformations $\Omega\to\lambda\Omega$, $\Omega\to\Omega^{1+a}$ and identifies $c$ and $d$ from the following limiting ratios: 
\begin{equation}    \lim_{\Omega\to\infty} \frac{S^{(h,g)}_{c,d}[(p(G_1),....,p(G_{\lambda\Omega}))]}{S^{(h,g)}_{c,d}[(p(G_1),....,p(G_{\Omega})]} = \lambda^{1-c},
    \label{eq:HT1}
\end{equation}
\begin{equation}  \lim_{\Omega\to\infty} \frac{S^{(h,g)}_{c,d}[(p(G_1),....,p(G_{\Omega^{1+a}})]}{S^{(h,g)}_{c,d}[(p(G_1),....,p(G_{\Omega}))]} \Omega^{a(c-1)}= (1+a)^d.
    \label{eq:HT2}
\end{equation}
Different choices of $h$ and $g$ may result in the same values of the entropic parameters, in which case the corresponding entropies are considered asymptotically equivalent~\cite{thurner}.
Therefore in this case the entropic parameters identify equivalence classes of entropies with the same asymptotic properties.
We will call the entropies that respect \emph{SK1-SK3}, plus Eq.~\eqref{composable}, the Hanel-Thurner (\emph{HT}) entropies.

\subsection{How to identify the correct entropy?\label{howto}}
On \emph{UJK} entropies, \emph{HT} entropies and in principle any generalized entropy family, it is important to ensure that the MEP can be reformulated consistently as a tool to construct probability distributions starting from observations of the system.
This procedure is sometimes called the Generalized Maximum-Entropy Principle (GMEP).
However, a number of serious conceptual and practical problems are currently open.\\

First, while it is still possible, for a fixed value of the entropic parameter(s),
to identify the functional form of the probability distribution maximizing the generalized entropy under certain `soft' constraints, it is no longer guaranteed in general that the enforcement of these constraints remains consistent with the application of the ML principle to the Lagrange multipliers and that the entropy retains a role for model selection as in Eq.~\eqref{Shannonlikelihood}.
Only for certain generalized entropies this consistency is retrieved, but not for all of them, as we show later with some notable examples.
Since the ML principle is agnostic with regard to the form of the probability distribution, and even more so to the type of entropy the latter maximizes, this inconsistency raises suspicion. Unfortunately, its possible origin is poorly discussed in the literature.\\

Second, fundamental problems arise when considering the determination of the entropic parameters themselves, or in other words of the `correct' entropy in a parametric family.
In particular, two main approaches have been proposed. 
One approach requires \emph{a priori} knowledge of the system (e.g. how certain properties of the entropy or of the system change with the number of accessible configurations~\cite{hanel2011comprehensive,thurner,balogh2020generalized}) as in Eqs.~\eqref{eq:HT1} and~\eqref{eq:HT2},
implying that, in absence of such knowledge, the entropic parameters cannot be consistently derived purely from data as the other parameters. 
Another approach does allow for the entropic parameters to be inferred from data, again invoking some form of maximization of the generalized entropy \cite{plastino2004general, bashkirov2004maximum}.
However, as we show below, this requirement conflicts with the ML principle, if the latter is extended to the estimation of the entropic parameters themselves.\\

Finally, when there are multiple i.i.d. observations available about the system (for instance in the case of a controlled repeated experiment), the generalized entropy should become somehow `consistent' also with Shannon entropy, because in the case of independence the axiom \emph{SJ3} should indeed lead to Shannon entropy as originally interpreted by Shore and Johnson.
So how can the same system be described by a non-Shannonian entropy when there is a single observation available and by a Shannonian entropy when there are multiple observations available? To the best of our knowledge, this puzzle is not discussed in the literature.

\section{One axiom to rule them all\label{results}}
The above limitations make the GMEP either inapplicable in practice without prior knowledge of the correct entropic parameter(s), or inconsistent with the ML principle and the information-theoretic consequences of \emph{SJ3} under independence. 
In the rest of this section, which contains all our results, we show that a possible solution to this problem can be achieved starting from a seemingly different viewpoint, i.e. by imposing an additional axiom that somehow `aligns' all entropies in a given family and therefore allows to select the most likely member of the family purely from data (if the latter contain information) and without prior knowledge of the system's properties.
Remarkably, the introduction of this simple requirement solves all the inconsistencies discussed in Sec.~\ref{howto}.

\subsection{The uninformativeness axiom}
We now introduce the axiom.
Unlike the \emph{SK} or \emph{SJ} ones, this axiom applies not to an individual entropy in a  generalized parametric family, but rather to the entire family.
Indeed the axiom does not represent yet another generalization of the \emph{SK} or \emph{SJ} ones, but rather an  `auxiliary' requirement to be added precisely when any such generalization is made, to restrict the form of the resulting entropic family.
\begin{itemize}
    \item \textit{Uninformativeness Axiom}: In a parametric family of entropies, the value of the entropy attained by the uniform distribution $P_u$ should not depend on the value of the entropic parameter(s).
\end{itemize}
Clearly, if the axiom is applied to families that include Shannon entropy $S_1$ as a particular case, it implies that all members of the family attain the same value $S_1[P_u]=\ln\Omega$ when applied to $P_u$.
This requirement equips generalized entropies with a universal scale and meaning.
As we show below, our axiom provides certain guarantees when the inference procedure is extended to the identification of the entropic parameters themselves. 
On one hand, the axiom ensures that no entropic parameter can be inferred from a completely uninformative (i.e. uniform) distribution, irrespective of how the parameter estimation procedure is conceived. 
On the other hand, when informative (non-uniform) data are available, the axiom ensures consistency with a generalized ML principle and model selection approach where all parameters, including the entropic one, can be identified from empirical observations, without prior knowledge of the system.\\

Note that, as required by \emph{SK2} and \emph{SJ5}, for a given value of $q$ the Uffink functional in Eq.~\eqref{uffink} is maximized by $P_u$.
This requirement comes from a `horizontal' perspective, in the sense that it holds for each $q$-entropy in the family. 
Our axiom, on the other hand, provides a `vertical' perspective: among all the $q$-entropies, none of them has to be preferred when applied to $P_u$. In other words, the axiom ensures the uninformativeness role of the uniform distribution not only for a specific entropy in the family, but across all of them. 
Since \emph{SK2} and \emph{SJ5} ensure that no entropy can exceed the value it attains on $P_u$, the axiom establishes a sort of common reference frame or universal scale, which allows to compare different entropies in a parametric family consistently. In particular, it ensures that all entropies in a parametric family that respects \emph{SK2} or \emph{SJ5} and includes Shannon entropy as a particular case attain values in the same interval $[0,\ln\Omega]$, irrespective of the value of the entropic parameter(s).
We will show that this guarantee ensures that the entropic parameter(s) can be estimated via a model selection approach purely from the input data, if the latter are informative (non-uniformly distributed).

\subsection{Application to important entropy families}

We now discuss some consequences of imposing the uninformativeness axiom to popular entropy families.\\

We start with the \emph{UJK} entropies $S^{(f)}_q[P]$ under the requirement that the family should include Shannon entropy as a particular case.
The entropy $S^{(f)}_q[P]=f(U_q[P])$, when evaluated on the uniform probability distribution $P_u=(\Omega^{-1},\dots,\Omega^{-1})$, returns the value
\begin{equation}
S^{(f)}_q[P_u]=f\left(U_q[P_u]\right) = f(\Omega)\qquad \textrm{for}\quad q \neq 1.
\label{neq1}
\end{equation}
Our axiom requires that $S^{(f)}_q[P_u]$ is independent of $q$, which implies that $f$ should be independent of $q$.
For $q=1$, technically $S^{(f)}_q[P]$ is only defined as the limit 
\begin{equation}
\lim_{q\to 1}S^{(f)}_q[P]=f\left(\lim_{q\to 1}U_q[P]\right),
\end{equation}
where we have used the $q$-independence of $f$.
If we require that, when $P=P_u$, this limit coincides with what Shannon entropy returns on $P_u$, i.e. $S_1[P_u]=\ln \Omega$, then we need a function $f$ such that 
\begin{equation}
\lim_{q\to 1}S^{(f)}_q[P_u]=f\left(\lim_{q\to 1}U_q[P_u]\right)=\ln\Omega,
\label{qlimit}
\end{equation}
i.e. $f(x)=\ln x$.
Therefore, combining Eqs.~\eqref{neq1} and~\eqref{qlimit}
we obtain $f(x)=\ln x$ for all $q$, i.e. the only viable \emph{UJK} entropy is R\'enyi entropy~\cite{renyi1961measures}
\begin{equation}
S_q[P]\equiv S^{(\ln)}_q[P]=    \ln U_q[P] =\frac{1}{1-q} \ln \sum_{i=1}^\Omega p^q(G_i),
\end{equation} 
where, since the entropy above is the only `surviving one' in the family $S^{(f)}_q[P]$, we have removed the superscript from the resulting $S^{(\ln)}_q[P]$.
From Eq.~\eqref{wShannon} we can confirm that this entropy reduces to Shannon entropy in the limit $q\to 1$, a well-known result for R\'enyi entropy.
This entropy is such that, on the uniform distribution $P_u$,
\begin{equation}
S_q[P_u]=\ln \Omega,
\label{unif}
\end{equation}
which does not depend on $q$, as demanded by our axiom. Therefore \emph{the only viable UJK entropy is R\'enyi entropy}. In general, other \emph{UJK} entropies do not respect our axiom.\\

An important counterexample is Tsallis entropy~\cite{tsallis1988possible}, defined as
\begin{equation}
S_q^\textrm{Tsallis}[P]\equiv S^{(\ln_q)}_q[P]=\frac{1}{1-q} \bigg(\sum_{i=1}^\Omega p^q(G_i)-1\bigg)
\label{tsallis}
\end{equation} 
and obtained from the so-called `$q$-logarithm' $f(x)= \ln_q (x) \equiv (x^{1-q}-1)/(1-q)$ (not to be confused with the ordinary logarithm of $x$ to base $q$): indeed, when evaluated on $P_u$, this entropy takes the $q$-dependent value
\begin{equation}
S_q^\textrm{Tsallis}\left[P_u\right]=\frac{\Omega^{1-q}-1}{1-q}=\ln_q (\Omega).
\label{tsallisuniform}
\end{equation}
From the point of view of our axiom, such $q$-dependence is a contradiction: different values of $q$ should not artificially attach different degrees of informativeness to an intrinsically uninformative distribution.
Seen from another point of view, this contradiction arises from the $q$-dependence of the function $f$ defining Tsallis entropy from the Uffink functional $U_q[P]$: such $q$-dependence is not admitted by our axiom because $f(U_q[P_u])$ should not depend on $q$. 
Note that the $q$-independence of the function $f$ defining the \emph{UJK} entropy $f(U_q[P_u])$ is a nontrivial consequence of our axiom, as it arises as necessary only when comparing entropies obtained for different values of $q$ (if only a single value of $q$ were considered, nothing would prevent $f$ from being specified by that value of $q$). In particular, our axiom would demand $q=1$ in order to have $S_q^\textrm{Tsallis}\left[P_u\right]=S_1\left[P_u\right]$, i.e. \emph{the only viable Tsallis entropy is Shannon entropy}.
We should stress at this point that the inadmissibility of Tsallis entropy is not in contradiction with the successful applications of the \emph{distribution} maximizing Tsallis entropy for fixed $q$~\cite{tsallis2009introduction}, because such distribution is exactly the same as the one maximizing R\'enyi entropy or any other monotonic function of the Uffink functional, as we also discuss later in this paper. However, when that distribution is `put back' into the entropy, only R\'enyi entropy gives consistent results in terms of the absolute quantification of the uncertainty and the associated ML estimation and model selection procedures. 
Indeed, as we show below, a ranking of models (or values of $q$) based on Tsallis entropy would `mess up' the ranking based on ML, while the use of R\'enyi entropy restores and extends the consistency with the ML principle.\\

As another example, we apply the uninformativeness axiom to the \emph{HT} family of composable $(c,d)$-entropies that can be written as in Eq.~\eqref{composable}.
If we require $S^{(h,g)}_{c,d}[P_u]=S_1[P_u]=\ln\Omega$ in analogy with Eq.~\eqref{unif}, then the axiom translates Eqs. (\ref{eq:HT1}) and (\ref{eq:HT2}) to:
\begin{equation}
    \lim_{\Omega\to\infty}\frac{\ln\lambda\Omega}{\ln\Omega}=\lambda^{1-c}
\end{equation}
\begin{equation}
    \lim_{\Omega\to\infty}\frac{\ln\Omega^{1+a}}{\ln\Omega}=(1+a)^d
\end{equation}
and implies $(c,d)=(1,1)$. 
This parameter choice identifies the equivalence class of entropies that are additive for independent events.
Both Shannon and R\'enyi entropies belong to this class.
In particular, in the case $h(x)=x$ (trace-form entropy) and $g(x)=-x\ln x$ (Shannon entropy), one gets $(c,d)=(1,1)$~\cite{thurner}, i.e. $S^{(x,-x\ln x)}_{1,1}[P]=S_1[P]$. Therefore \emph{Shannon entropy is a viable trace-form HT entropy} under our axiom.
Similarly, in the case $h(x)=\ln (x)/(1-q)$ and $g(x)=x^q$ (R\'enyi entropy) one again gets $(c,d)=(1,1)$~\cite{thurner}, i.e. $S^{(\ln (x)/(1-q),x^q)}_{1,1}[P]=S_q[P]$.
Therefore \emph{R\'enyi entropy is a viable composable HT entropy}.
By contrast, the case $h(x)=x$ and $g(x)=(x^q-\Omega^{-1})/(1-q)$ (Tsallis entropy) leads to $(c,d)=(q,0)$~\cite{thurner}, confirming that Tsallis entropy (which is another trace-form entropy) does not respect our axiom.\\

The fact that, for both the \emph{UJK} and \emph{HT} families, only R\'enyi entropy (or an asymptotically equivalent one) `survives' our axiom does not disagree with the possibility of non-extensivity of the entropy, which has led to the introduction of many variants of entropy over the last decades~\cite{tsallis2009introduction,thurner}.
Indeed, while our axiom selects entropy additivity for independent systems (as both Shannon and R\'enyi do), it does not have direct implications when independence is not present or even not known.
In particular, it should be stressed that non-extensivity is a property not of the entropy itself, but of how the number $\Omega$ of configurations scales with the physical size of the system (i.e. the number $n$ of units or particles)~\cite{thurner}. 
Even Shannon entropy can be non-additive if applied to a system where $\Omega$ (or $\Omega_{C^*}$, when in presence of a constraint $C^*$) is not exponential in $n$, as clear from Eq.~\eqref{boltzmann} or~\eqref{unif2}.
Note that Eq.~\eqref{unif2} applies in the microcanonical case, but a similar non-extensive scaling of the entropy would be exhibited in the canonical case as well. An important example in this respect is provided by random graphs: the number of all binary graphs on $n$ vertices is $\Omega=2^{\binom{n}{2}}$, so it is super-exponential~\cite{thurner,qi}.
Even when subject to various types of constraints $C^*$, the number $\Omega_{C^*}$ remains super-exponential, yet Shannon entropy is an appropriate entropy for random graph ensembles~\cite{squartini2017maximum}.
At the opposite extreme, even for systems where $\Omega$ does increase exponentially in $n$, the system may still be subject to certain constraints such that $\Omega_{C^*}$ is sub-exponential in $n$, so that the resulting entropy is sub-extensive. An example is the class of State Space Reducing processes~\cite{hanel2011comprehensive}.
Therefore one first general result implied by the uninformativeness axiom is that non-extensivity or non-ergodicity (when present) should be completely encoded in the scaling of $\Omega_{C^*}$ with $n$, thus ultimately in the identification of the proper (effective) constraint $C^*$, and not in the expression of the entropy itself.

\subsection{The generalized MEP}
In a GMEP context, a direct consequence of the fact that our axiom restricts the viable expressions for the generalized entropies is, of course, a corresponding restriction on the probability distributions maximizing such generalized entropies under soft constraints (note that, under hard constraints, all maximum-entropy distributions reduce to the microcanonical uniform distribution $P_0$ described in Sec.~\ref{MEP}). 
This restriction can have two (related) effects: one on the functional form of the maximum-entropy distribution and one on the way  the distribution connects to the entropy itself and possibly other quantities. The \emph{HT} and \emph{UJK} entropies serve as good examples for both effects, as we now show.\\

For instance, while the general form for the probability distribution that maximizes the \emph{HT} entropy $S^{(h,g)}_{c,d}[P]$ in trace form ($h(x)=x$) is the exponential of the so-called Lambert-W function\footnote{The Lambert-W function $\mathcal{W}(x)$, which cannot be written in close form, is the solution to the equation $x=\mathcal{W}(x)e^{\mathcal{W}(x)}$. The real solutions are those that are relevant here.}  $\mathcal{W}(x)$~\cite{hanel2011comprehensive,thurner}, the only admissible form according to our axiom is the one corresponding to the choice $(c,d)=(1,1)$.
With this parameter choice, the $\mathcal{W}(x)$ function reduces to a linear function, so that the maximum-entropy probability reduces to the Boltzmann-Gibbs distribution in Eq.~\eqref{BG}~\cite{thurner}, consistently with the fact that the only admissible trace-form \emph{HT} entropy according to our axiom is Shannon entropy, as we have shown above.
To obtain a truly generalized maximum-entropy probability, one should therefore consider non-trace-form entropies.\\

In particular, considering the R\'enyi entropy $S_q[P]$ which our axiom selects from both the \emph{UJK} and the \emph{HT} families, the GMEP can be formulated as the following well-known generalization of the MEP described in Sec.~\ref{MEP}.
Given an empirically observed value $C^*$ of a (scalar or vector) function $C(G)$ of the unknown microstate $G$ of a system, the least biased inference about $G$ is provided by the distribution $P_q$ that maximizes $S_q[P]$ under the (soft) constraint
\begin{equation}
\langle C \rangle_q \equiv \frac{\sum_{i=1}^\Omega p^q(G_i)\,C(G_i) }{\sum_{i=1}^\Omega p^q(G_i)}=C^*,
\label{qavg}
\end{equation}
which generalizes the usual Shannonian constraint in Eq.~\eqref{mean} (note that $\langle C \rangle_1=\langle C \rangle$).
The quantity
$\langle C \rangle_q$ is sometimes called \emph{(normalized) $q$-mean}, and it can be regarded as a mean with respect to the so-called \emph{escort} (or \emph{zooming}) probability distribution $\tilde{p}(G_i)=p^q(G_i)/\sum_{j} p^q(G_j)$ \cite{beck1995thermodynamics,tsallis2009introduction}. 
This $q$-mean has been introduced to extend important properties and relations from the classical (i.e. Shannonian) statistical mechanics to the non-extensive one, including the Legendre structure of thermodynamics, the $H$-theorem and the Ehrenfest theorem \cite{tsallis2009introduction}. However, from the point of view of statistical inference, it has always been debated whether or not $\langle C \rangle_q$ is a proper constraint, since it lacks a direct interpretation in relation to the available data. 
Here, we choose the $q$-mean for a reason that is both conceptual and pragmatic: it ensures that the expected value of the constraint is always finite as soon as the distribution is normalizable (including cases when the ordinary mean $\langle C \rangle$ diverges, namely when $q>3/2$) and moreover it remains consistent with the ML principle, as we show later on.
Both requirements are natural in our setting (described later) where we want to be able to determine $q$ purely from the data without prior knowledge of its value and therefore without knowing whether the ordinary mean would diverge.\\

To carry out the constrained maximization of $S_q[P]$, we look for the vanishing derivatives of the $q$-Lagrangian
\begin{equation}
\label{gmepq}
\mathcal{L}_q[P]=S_q[P]-\alpha\left[\sum_{i=1}^\Omega p(G_i)-1\right]-\theta\cdot\left[\langle C \rangle_q - C^*\right]
\end{equation}
with respect to $P$, $\alpha$ and $\theta$, and assume $q\ne 1$ from now on. 
The resulting values are denoted as $P_q,\alpha_q,\theta_q$. 
In particular, setting $\partial\mathcal{L}_q[P]/\partial P|_{P_q}=0$ we get
\begin{eqnarray}   0&=&\left.\frac{\partial\mathcal{L}_q[P]}{\partial p(G_i)}\right|_{p_q(G_i)} \label{eq:maxL}\\
    &=&\frac{q}{1-q}\frac{p_q^{q-1}(G_i)}{\sum_j p_q^q(G_j)}-\alpha- q\,p_q^{q-1}(G_i)\frac{\theta\cdot(C(G_i)-\langle C \rangle_q)}{\sum_j p_q^q(G_j)}\nonumber 
\end{eqnarray}
for all $i$ from 1 to $\Omega$, from which it is clear that $p_q(G_i)$ depends on $\theta$, as in the case $q=1$, and additionally on $q$.
The derivative of $\mathcal{L}_q[P]$ with respect to $\alpha$ leads to a condition identical to Eq.~\eqref{alpha1}:
\begin{equation}
\left.\frac{\partial \mathcal{L}_q[P_q]}{\partial \alpha}\right|_{\alpha_q}=0\quad\Rightarrow\quad\sum_{i=1}^\Omega p_q(G_i,\theta)\,=1,
\label{alphaq}
\end{equation}
which can be used to determine $\alpha_q$ by multiplying both sides of Eq.~\eqref{eq:maxL} and then summing over $i$. 
We then get
\begin{equation}
    \alpha_q=\frac{q}{1-q}\qquad (q\ne 1),
    \label{alphaqbis}
\end{equation}
which is the counterpart of Eq.~\eqref{alpha1bis}. Substituting $\alpha_q$ in (\ref{eq:maxL}) and singling out $p_q(G_i)$ yields
\begin{equation}
p_q(G_i,\theta)=\frac{[1-(1-q)\,\theta\cdot(C(G_i)-\langle C \rangle_q)]_+^{{1}/{(1-q)}}}{\left[\sum_{j=1}^\Omega p_q^q(G_j,\theta)\right]^{{1}/{(1-q)}}}
\label{preq}
\end{equation}
where we have used the notation $[x]_+^a\equiv0$ if $x<0$, while $[x]_+^a\equiv x^a$ otherwise~\cite{tsallis2009introduction}.
Note that the denominator of Eq.~\eqref{preq} equals the Uffink functional $U_q[P_q(\theta)]$ and  must also equal  the \emph{generalized partition function}
\begin{equation}
W_q(\theta)\equiv\sum_{i=1}^\Omega [1-(1-q)\,\theta\cdot(C(G_i)- \langle C \rangle_q)]_+^{1/(1-q)}
\end{equation}
since $p_q(G_i,\theta)$ is already normalized via the condition in Eq.~\eqref{alphaqbis}.
In other words,
\begin{equation}
    \label{eq:WtoU}
    W_q(\theta)=\left[\sum_{i=1}^\Omega p_q^q(G_i,\theta)\right]^{1/(1-q)}=U_q[P_q(\theta)].
\end{equation}
Finally, the maximum-entropy probability equals
\begin{equation}
 p_q(G_i,\theta)=\frac{\left[1-(1-q)\,\theta\cdot(C(G_i)- \langle C \rangle_q)\right]_+^{1/(1-q)}}{W_q(\theta)}
 \label{qexp}
\end{equation}
which has the form of a so-called $q$-exponential~\cite{tsallis2009introduction} distribution.
Note that Eqs.~\eqref{gmepq} and~\eqref{qexp} generalize Eqs.~\eqref{mepq} and~\eqref{BG}, respectively.
Moreover note that, if we formally introduce a pseudostate $\tilde{G}$ such that $C(\tilde{G})=\langle C \rangle_q$, it follows from Eq. (\ref{qexp}) that $p_q(\tilde{G},\theta)= 1/W_q(\theta)=1/U_q[P_q(\theta)]$. Then, from Eq. (\ref{eq:WtoU}), one can see that:
\begin{equation}
    p^{q-1}_q(\tilde{G},\theta) = \sum_{i=1}^\Omega p_q^q(G_i,\theta)=U_q^{1-q}[P(\theta)].
    \label{eq:23}
\end{equation}
We will discuss the relationship between $C(\tilde{G})$ and $C(G^*)$ later.\\

When $q\to 1$, $p_q(G_i,\theta) \rightarrow {Z^{-1}_1(\theta)}\exp(-\theta\cdot C(G_i))$, retrieving the Boltzmann-Gibbs distribution in Eq.~\eqref{BG}.
When $q\ne 1$, the $q$-exponential has nothing to do with the ordinary exponential and actually has power-law tails proportional to $C(G_i)^{1/(1-q)}$ for large values of $C(G_i)$.
The presence of these heavy tails, which are widespread in several real-world complex systems, is one of the reasons why $q$-exponentials have attracted interest, their derivation from the maximization of a suitable entropy appearing convenient and parsimonious~\cite{thurner,tsallis2009introduction}.
In the literature, there is some emphasis on the fact that $q$-exponentials derive from the maximization of Tsallis entropy given by Eq.~\eqref{tsallis}. However, they rather derive from \emph{any} of the \emph{UJK} entropies in Eq.~\eqref{uffink}: the distribution maximizing $U_q[P]$ necessarily maximizes $f(U_q[P])$ as well, for any monotonic $f$. 
Indeed, our derivation above started from R\'enyi entropy and is also already well known.
The real differences among the members of the \emph{UJK} entropy family arise when the maximum-entropy $q$-exponential is put back into the entropy itself. When this happens, the uninformativeness axiom has the important role of selecting R\'enyi entropy as the member of the family that solves all the inconsistencies discussed in Sec.~\ref{howto}, as we show later in the paper.\\

What remains to be done is the determination of the parameter $\theta$. It is useful at this point to introduce the reparameterization \begin{equation}
    \psi(\theta) \equiv \frac{\theta}{1+(1-q)\,\theta \cdot\langle C \rangle_q},
\end{equation} through which it is possible to (formally) remove $\langle C\rangle_q$ from the expression for $p_q(G_i,\theta)$ and get
\begin{equation}
 \label{pqavg}
p_q(G_i,\psi)=\frac{\left[1-(1-q)\,\psi\cdot C(G_i)\right]_+^{1/(1-q)}}{Z_q(\psi)}
\end{equation}
where, denoting the inverse of $\psi(\theta)$ as $\theta(\psi)$,
\begin{eqnarray}
Z_q(\psi)&\equiv&\sum_{i=1}^\Omega[1-(1-q)\,\psi\cdot C(G_i)]_+^{1/(1-q)}\\
&=&\frac{W_q\left(\theta(\psi)\right)}{[1+(1-q)\,\theta(\psi)\cdot\langle C \rangle_q]^{1/(1-q)}}
\end{eqnarray}
is the reparametrized partition function. 
Note that $Z_q(\psi)\neq W_q\left(\theta(\psi)\right)$ unless $q\to 1$, in which case $\psi\to\theta$ and $W_1(\theta)\to Z_1(\theta)$.
The optimal value $\psi_q$ is determined by the condition 
\begin{equation}
\left.\frac{\partial \mathcal{L}[P_q]}{\partial \psi}\right|_{\psi_q}=0\quad\Rightarrow\quad\frac{\sum_{i=1}^\Omega p^q_q(G_i,\psi_q)\,C(G_i)}{\sum_{i=1}^\Omega p^q_q(G_i,\psi_q)}=C^*
\label{psiq}
\end{equation}
corresponding to the intended requirement in Eq.~\eqref{qavg} and generalizing Eq.~\eqref{theta1} to the case $q\ne 1$.
One the value $\psi_q$ is determined via the condition above, it can be inserted into Eq.~\eqref{pqavg} to obtain the final maximum-entropy probability distribution $P_q(\psi_q)$.\\

As a final remark here, we note that if one constrains the ordinary mean $\langle C \rangle$ rather than the $q$-mean $\langle C \rangle_q$ and follows the same maximization procedure for $S_q[P]$ as described above, a different maximum-entropy distribution $\hat{P}_q(\theta)$ is obtained:
\begin{equation}
    \hat{p}_q(G_i)=\frac{[1-(q-1)\,\hat{\theta}\cdot (C(G_i)-\langle C \rangle)]_+^{\frac{1}{q-1}}}{\hat{W}_q(\hat{\theta})}.
\end{equation}
Following the reparameterization previously introduced, it is also possible to write:
\begin{equation}
    \hat{p}_q(G_i,\hat{\psi})=\frac{\left[1-(q-1)\,\hat{\psi}\cdot C(G_i)\right]_+^{1/(q-1)}}{\hat{Z}_q(\hat{\psi})},
\end{equation}
where
\begin{equation}
    \hat{\psi}(\hat{\theta}) \equiv \frac{\hat{\theta}}{1+(q-1)\,\hat{\theta} \cdot\langle C \rangle}.
\end{equation}
Note that the transformation $q \to 2-q$ formally links the two types of constraint. 
In particular, one can see that
\begin{equation}
   \hat{p}_q(G_i, \hat{\psi})= p_{2-q}(G_i, \hat{\psi}).
\end{equation}

\subsection{Link with the ML principle and model selection\label{newML}}
We now show that the entropy  selected by the uninformativeness axiom restores consistency with the ML principle and retains an interpretation for model selection, exactly as in the Shannon case. Both properties are not guaranteed for other entropies.
At the same time, we show how to account for multiple independent observations about the same system.

In analogy with Sec.~\ref{sec:ML}, we start with the case $M=1$ and 
define the ML estimation procedure for the parameter $\psi_q$ as follows:
\begin{equation}
\psi_q^*=\argmax_\psi {\ell_q(\psi)},\quad 
\ell_q(\psi)\equiv\ln p_q(G^*,\psi).
\label{MLq}
\end{equation}
Requiring $\partial{\ell_q(\psi)}/\partial{\psi}|_{\psi_q^*}=0$, one gets
\begin{equation}
    \sum_{i=1}^\Omega C(G_i)\,p_q^{q}(G_i,\psi^*_q)= C(G^*)\,p_q^{q-1}(G^*, \psi^*_q)
\end{equation}
and, dividing both terms by $\sum_G p_q^q(G^*, \psi^*_q)$,
\begin{equation}
    \langle C \rangle_q = \frac{C(G^*)\,p_q^{q-1}(G^*, \psi^*_q)}{\sum_G p_q^q(G^*, \psi^*_q)}.
    \label{eq:22}
\end{equation}
One might think that the right hand side of the above equation is different from the `desired' value $C^*=C(G^*)$, however this is not the case.
Indeed, considering again a pseudostate $\tilde{G}$ such that $C(\tilde{G})=\langle C\rangle_q$ and using Eq.~\eqref{eq:23}, we can rewrite Eq.~\eqref{eq:22} as
\begin{equation}
    \frac{C(\tilde{G})}{C(G^*)}=\frac{1-(1-q)\,\psi_q^*\cdot C(\tilde{G})}{1-(1-q)\,\psi_q^*\cdot C(G^*)},
\end{equation}
which leads to $C(\tilde{G})=C(G^*)$.
In other words, the value $\psi^*_q$ defined by Eq.~\eqref{MLq} coincides with the value $\psi_q$ defined by Eq.~\eqref{psiq}, i.e.  $\psi_q^*\equiv\psi_q$, i.e.
\begin{equation}
\left.\frac{\partial \ell_q(\psi)}{\partial \psi}\right|_{\psi^*_q}=0\quad\Rightarrow\quad\frac{\sum_{i=1}^\Omega p^q_q(G_i,\psi^*_q)\,C(G_i)}{\sum_{i=1}^\Omega p^q_q(G_i,\psi^*_q)}=C^*
\label{psiq*}
\end{equation}
in analogy with Eq.~\eqref{psiq}.
This means that the ML principle can still be seen as equivalent to the part of the Lagrangian optimization relative to $\psi$.
Moreover, the application of the logarithm to both sides of Eq. (\ref{eq:23}) leads to
\begin{equation}
    \ell_q(\psi^*_q)=-S_q[P_q(\psi^*_q)],
    \label{Renyilikelihood}
\end{equation}
showing that, for $M=1$, the log-likelihood of the observation coincides with minus the R\'enyi entropy.
This extends Eq. (\ref{Shannonlikelihood}) to the case $q\ne 1$, generalizing the result that, in a model selection framework, different models can be ranked according to their maximized likelihood or, equivalently, to their realized R\'enyi entropy. Notably, other entropies of the \emph{UJK} family, including Tsallis entropy, do not manifest this property.
Also the relationship in Eq.~\eqref{lagranlike} generalizes as follows:
\begin{equation}
\mathcal{L}_q[P_q(\psi_q^*)]=S_q[P_q(\psi_q^*)]=-\ell_q(\psi_q^*),
\label{lagranlikeq}
\end{equation}
relating the value of the Lagrangian attained by $P_q(\psi_q^*)$ to the maximized log-likelihood.
Therefore, up to this point, it seems that R\'enyi entropy retains all the desirable properties of Shannon entropies.\\

We now consider the case of $M>1$ i.i.d. realizations $\{G_m^*\}_{m=1}^M$ of the system, leading to $M$ independent observations $\{C_m^*\}_{m=1}^M$ of the constraint, where $C_m^*\equiv C(G^*_m)$ for all $m$.
We have already seen in Sec.~\ref{sec:ML} that in this case it is the ML principle, not the MEP, that identifies how to combine the $M$ observed values. 
Introducing again the average log-likelihood $\overline{\ell}_q(\psi)$, the ML condition for $\psi$ becomes a straightforward generalization of Eq.~\eqref{MLM}:
\begin{equation}
\psi_q^*=\argmax_\psi {\overline{\ell}_q(\psi)},\quad  
\overline{\ell}_q(\psi)\equiv\frac{\sum_{m=1}^M\ln p_q(G_m^*,\psi)}{M}.
\label{MLMq}
\end{equation}
It is not difficult to show that requiring $\partial \overline{\ell}_q(\psi) /\partial\psi|_{\psi^*_q}=0$ translates into:
\begin{equation}
     \sum_{i=1}^\Omega C(G_i)\,p_q^{q}(G_i,\psi^*_q)=\frac{1}{M}\sum_{m=1}^M C(G^*_m)\,p_q^{q-1}(G^*_m, \psi^*_q)
     \label{MLCq}
\end{equation}
or equivalently
\begin{equation}
     \langle C\rangle_q=\frac{\sum_{m=1}^M C(G^*_m)\,p_q^{q-1}(G^*_m, \psi^*_q)}{M\sum_{i=1}^\Omega p_q^{q}(G_i,\psi^*_q)}
     \label{MLCq2}
\end{equation}
which extends the classical  ($q=1$) result in Eq.~\eqref{arithm} to the general, non-Shannon case. 
We therefore learn that the arithmetic average is no longer the optimal way of combining the $M$ available observations in order to determine the parameter $\psi$.
Indeed, dismissing the arithmetic average makes sense if we recall that, \emph{a priori}, we do not even know whether the first moment of the distribution generating the $M$ values $\langle C^*_m\}_{m=1}^M$ is finite.
Indeed, the $q$-exponential distributions that are solution to the GMEP exhibit a power-law behavior for $q\ne 1$. 
As a consequence, in principle all their moments could diverge, depending on the value of $q$. Assuming that $q$ is not known beforehand and is rather determined by the inference procedure itself (as we assume later on), it would make no sense at all to use the arithmetic average to constrain the $q$-mean in case of multiple observations, since that average might become infinite in the $M\to\infty$ limit when $q>3/2$, while the $q$-mean is by construction finite whenever the distribution is normalizable. 
The same problem might in principle apply to any higher moment $\langle C^n \rangle$ with $n>1$, while any $q$-generalized moment $\langle C^n \rangle_q$ evaluated with respect to Eq.~(\ref{pqavg}) converges if $q$ is such that the distribution is normalizable (which is a basic requirement for this procedure to be consistent \cite{tsallis2009introduction}). 
The ML estimator determined by Eq.~\eqref{MLCq2} identifies the distribution's parameters, irrespective of the converge of any moment.\\

An important consequence of the fact that $\langle C\rangle_q$ is no longer equal to the arithmetic mean of the $M$ observations is that in general, for $q\ne 1$ and $M>1$,
\begin{equation}
S_q[P_q({\psi^*_q})]\ne-\overline{\ell}_q(\psi^*_q),
\label{RenyilikelihoodM}
\end{equation} 
thus failing to generalize Eq.~\eqref{ShannonlikelihoodM} to the case $q\ne 1$ and Eq.~\eqref{Renyilikelihood} to the case $M>1$.
Similarly, Eqs.~\eqref{lagranlike} and~\eqref{lagranlikeq} do not generalize here.
Rather, a relationship that is still valid is
\begin{equation}
S_q[P_q({\psi^*_q})]=-\tilde{\ell}_q(\psi^*_q),
\label{Renyipseudo}
\end{equation}
where $\tilde{\ell}_q(\psi)\equiv\ln p_q(\tilde{G},\psi)$ is a sort of `pseudolikelihood' involving the pseudostate $\tilde{G}$ such that $C(\tilde{G})=\langle C\rangle_q$ introduced above. Unfortunately, $\tilde{\ell}_q(\psi)$ is no longer equal to the actual log-likelihood $\overline{\ell}_q(\psi)$ based on the $M$ observations.
Does this mean that, in presence of multiple i.i.d. observations of the same quantity about a system, the correspondence between log-likelihood and entropy is lost?
The answer to this question emerges when looking at a seemingly unrelated problem, i.e. the selection of the optimal value of the entropic parameter $q$,  and is provided below.

\subsection{Inference of the entropic parameter}
%For instance, within a GMEP Eq.~\eqref{tsallisuniform} might suggest looking for the value of $q$ that maximizes $S_q^\textrm{Tsallis}$, {\color{red} an attempt that would always return the value ...}. 
%To prevent this from happening, in most approaches one generally treats the entropic parameters as special `external' parameters, i.e. specified \emph{ad hoc} and, unlike the parameters $\theta$, not obtainable from the data through an inference procedure.
%By contrast, by ensuring that different values of the entropic parameters return different values of the entropy only when these differences are supported by the data, our axiom allows to treat the entropic parameters like all other parameters and infer them from the data, as we now show.

We now come to the last, and in many ways most crucial, benefit implied by the uninformativeness axiom, namely the possibility of consistently identifying the entropic parameter(s) purely from the data, without postulating \emph{a priori} knowledge about the system — such as scaling laws of the type exemplified by Eqs.~\eqref{eq:HT1} and~\eqref{eq:HT2}~\cite{hanel2011comprehensive,thurner,balogh2020generalized}.

To this end, starting directly with the general case $M\ge 1$, we invoke again the ML principle and, building on its restored consistency with the estimation of the other parameters of the maximum-entropy distribution proven in Eq.~\eqref{psiq*}, extend it to the identification of the entropic parameter(s) themselves. Indeed, the ML principle treats any parameter agnostically, without specific interpretations, and is therefore `unaware’ of the fact that $q$ and the other structural parameters play different roles in an information-theoretic setting. 
Considering again Re\'nyi entropy as the only viable entropy from the \emph{UJK} and \emph{HT} families, the ML principle applied to the entropic parameter $q$ is formally stated as follows:
\begin{equation}
q^*=\argmax_q {\overline{\ell}_q(\psi)},\quad  
\overline{\ell}_q(\psi)\equiv\frac{\sum_{m=1}^M\ln p_q(G_m^*,\psi)}{M}.
\label{MLMqq}
\end{equation}
On the other hand, combining the above expression with Eq.~\eqref{MLMq}, it is clear that the estimation of $q$ is coupled to that of $\psi$, so that the actual formulation of the extended ML principle is
\begin{equation}
(\psi^*_{q^*},q^*)=\argmax_{(\psi,q)} {\overline{\ell}_q(\psi)},\quad  
\overline{\ell}_q(\psi)\equiv\frac{\sum_{m=1}^M\ln p_q(G_m^*,\psi)}{M}.
\label{MLMqqq}
\end{equation}
This expression immediately tells us that, once the ML principle is extended to the determination of $q$, the results we have discussed in Sec.~\ref{newML} represent only one side of the coin. 
Now, requiring jointly
\begin{equation}
\left.\frac{\partial \overline{\ell}_q(\psi)}{\partial\psi}\right|_{(\psi^*_{q^*},q^*)}=0,\quad 
\left.\frac{\partial \overline{\ell}_q(\psi)}{\partial q}\right|_{(\psi^*_{q^*},q^*)}=0,
\end{equation}
we arrive again at Eq.~\eqref{MLCq2} (with $q$ replaced by $q^*$) plus the additional condition
\begin{eqnarray}
\label{eq:maxq_1}
 &&\sum_{i=1}^\Omega p_{q^*}(G_i,\psi^*_{q^*})\ln\left[1-(1-q^*)\,\psi^*_{q^*}\cdot C(G_i)\right]\\
 &=&\frac{1}{M}\sum_{m=1}^M p_{q^*}(G^*_m,\psi^*_{q^*})\ln\left[1-(1-q^*)\,\psi^*_{q^*}\cdot C(G_m)\right].\nonumber
\end{eqnarray}
Recalling from Eq.~\eqref{pqavg} that
\begin{equation}
    1-(1-q^*)\,\psi^*_{q^*}\cdot C(G_i) = [p_{q^*}(G_i,\psi^*_{q^*})Z_{q^*}(\psi^*_{q^*})]^{1-q^*}
\end{equation}
we obtain the condition
\begin{equation}
 \sum_{i=1}^\Omega p_{q^*}(G_i,\psi^*_{q^*})\ln p_{q^*}(G_i,\psi^*_{q^*})=\frac{1}{M}\sum_{m=1}^M \ln p(G^*_m,\psi^*_{q^*}).
\end{equation}
In other words, the additional ML condition determining $q^*$ requires that \emph{the maximized log-likelihood equals minus Shannon entropy}, i.e.
\begin{equation}
    S_1[P_{q^*}(\psi^*_{q^*})]=-\overline{\ell}_{q^*}(\psi^*_{q^*}),
    \label{1oss_q}
\end{equation}
restoring an analogy with Eq.~\eqref{ShannonlikelihoodM} that appeared to be lost and replaced by Eq.~\eqref{Renyipseudo} when considering $q\ne 1$.
Actually, we now realize that, when the ML principle is extended to $q$, the correspondence with Eq.~\eqref{ShannonlikelihoodM} is not replaced, but rather \emph{accompanied} by Eq.~\eqref{Renyipseudo}.
Remarkably, the connection between Shannon entropy and log-likelihood at the specific parameter value $(\psi^*_{q^*},q^*)$ remains a general result, even for $q\ne 1$ and $M>1$. 
This might look quite surprising, because, for $q\ne 1$, the log-likelihood is based on the $q$-exponential distribution that maximizes R\'enyi, not Shannon, entropy.\\

Despite the surprise, the above result makes perfect sense because we have assumed $M$ independent observations.
Actually, it solves the final inconsistency we pointed out in Sec.~\ref{howto}: assuming independent observations justifies Shore and Johnson's original restricted interpretation of axiom \emph{SJ3} and leads to Shannon entropy as the quantifier of the uncertainty of the data.
Indeed the inequality in Eq.~\eqref{RenyilikelihoodM} should be put in relation with our initial discussion of the axiom \emph{SJ3} about system independence. Recall that assuming that the $M$ values $\{C^*_m\}_{m=1}^M$ come from independent observations is equivalent to assuming that there are $M$ identical and independent copies of the same system, each copy being observed exactly once. 
Under this assumption of independence, the original reasoning by Shore and Johnson becomes appropriate and one should therefore expect that Shannon entropy, rather than R\'enyi entropy, is the proper entropy describing the combined system of $M$ copies.
Therefore the breakdown of the correspondence between the average log-likelihood and R\'enyi entropy can be regarded as a symptom of the assumed independence of the $M$ observations.
When $M=1$, we can use Eq.~\eqref{Renyilikelihood} and combine it with Eq.~\eqref{1oss_q} to obtain
\begin{equation}
   S_{q^*}[P_{q^*}(\psi^*_{q^*})] = S_1[P_{q^*}(\psi^*_{q^*})] 
   \label{1oss_final}
\end{equation}
showing that in this particular case the maximum-entropy probability distribution returns coinciding values of Shannon and R\'enyi entropy, even if it maximizes the latter but not the former. 
This result does not in general for $M>1$.\\

The remarkable result in Eq.~\eqref{1oss_q} has an important consequence for model selection.
In particular, in order to determine both $q^*$ and $\psi^*_{q^*}$, one can consider a range of values for $q$ and, for each value in the range, compute $\psi^*_q$ according to Eq.~\eqref{MLCq2}.
This results, for each value of $q$ in a log-likelihood $\overline{\ell}_q(\psi^*_q)$ that is only partially maximized, in the sense that the maximization has been carried out only with respect to $\psi_q$, and not yet with respect to $q$. 
Then, among all these partially maximized log-likelihoods, one can select the one with the largest value. 
This will identify the value $q^*$ and the associated value $\psi^*_{q^*}$, which ultimately correspond to the completely maximized log-likelihood $\overline{\ell}_{q^*}(\psi^*_{q^*})$.
Only for this parameter choice $(q^*,\psi^*_{q^*})$, the log-likelihood equals minus Shannon entropy.
%Notice that condition (\ref{1oss_final}) is necessary but not sufficient for $(\psi^*_{q^*},q^*)$ to be the ML estimators. For example, the couple of values $(\psi^*_1,1)$ is always a solution of (\ref{1oss_final}), but it may not correspond to the ML estimator, as we will point out in a numerical example.\\
So from the ML condition Shannon entropy emerges spontaneously: while the probability $P_q$ maximizes R\'enyi entropy and not Shannon entropy, the latter is the correct entropy for model selection to take independence into account.
It follows that the introduction of our axiom leads to an entropy-grounded model selection criterion, based on the maximization of the R\'enyi entropy to obtain the functional form of the probability distribution and the ML principle to estimate its parameters, including the entropic one.
In order to illustrate the performance of the above approach, we now consider two simple numerical examples.\\

%%%%%%%%%%%%%%%%%%
\begin{figure*}[t]
\includegraphics[width=1.0\textwidth]{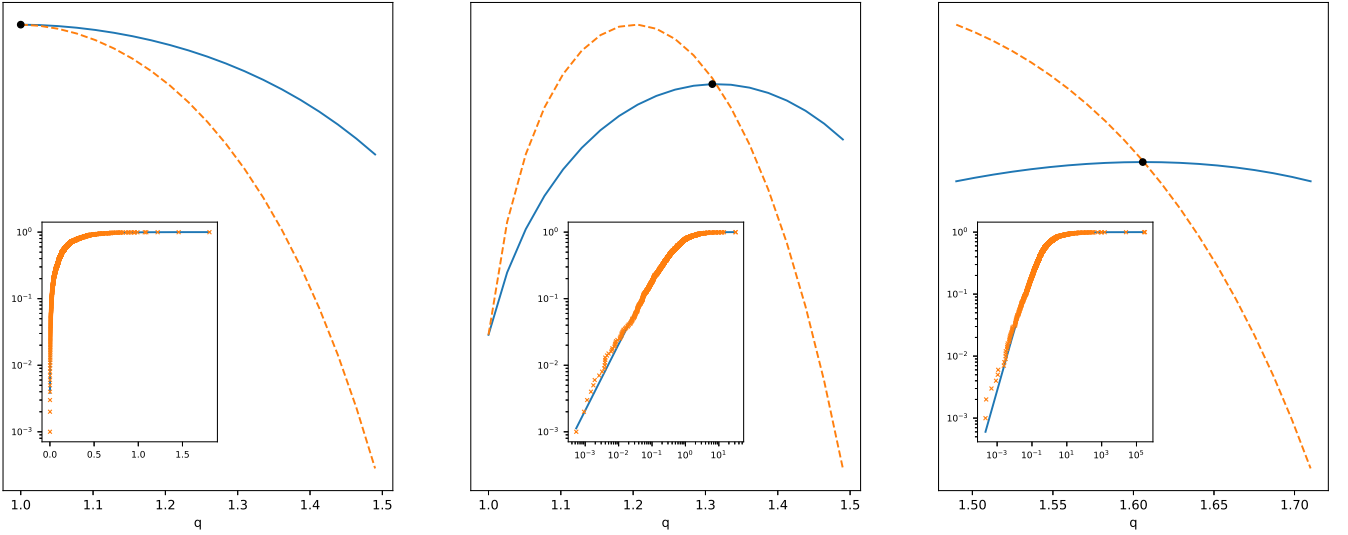}
\caption{\label{fig} Comparison between the average partially maximized log-likelihood $\overline{\ell}_q(\psi^*_q)$ (solid line) and minus Shannon entropy $-S_1[P_q(\psi^*_q)]$ (dashed line) as a function of $q$, for three samples of $M=10^3$ deviates generated from the probability distribution $P_q(\psi)$ in Eq.~\eqref{eq:pareto}, and in particular: exponential distribution where $q_{\textrm{true}}=1$ and $\psi_{\textrm{true}}=5.0$ (left), $q$-exponential (power-law) distribution with finite first moment where $q_{\textrm{true}}=1.3$ and $\psi_{\textrm{true}}=3.0$ (center), and $q$-exponential (power-law) distribution with diverging first moment where  $q_{\textrm{true}}=1.6$ and $\psi_{\textrm{true}}=7.0$ (right). The insets show the comparison between the empirical cumulative distributions of the $M$ realized values (crosses) and the retrieved maximum-entropy distribution using the inferred values $(q^*,\psi^*_{q^*})$ (solid line).}
\end{figure*}
%%%%%%%%%%%%%%%%%

Our first example is a system described by an observable $C(G)$ taking only positive real values, i.e. $C(G) \in [0, +\infty)$. Moreover, we assume that $\Omega_{C}=1$ for all $C$, meaning that for each value $C(G)$ of the observable there is only one state $G$ that realizes it. 
Thus, the sums over system states simplify into integrals over the observable values: $\sum_G \rightarrow \int_0^\infty dC(G)$. 
The probability distribution resulting from the GMEP is then:
\begin{equation}
\label{eq:pareto}
p_q(G_i,\psi)= (2-q)\,\psi\left[1-(1-q)\,\psi\cdot C(G_i)\right]_+^{\frac{1}{1-q}},
\end{equation}
where we have used $Z_q(\psi)=1/{(2-q)\psi}$.
For different values of $\psi$ and $q$, we have drawn an i.i.d. sample of $M=10^3$ realizations from the distribution above, with the aim of inferring the true value of those parameters purely from the data so generated. 
In particular, we have generated samples from an exponential distribution (i.e. $q_{\textrm{true}}=1$), a $q$-exponential distribution with finite first moment $\langle C\rangle$  ($q_{\textrm{true}}=1.3$) and a $q$-exponential distribution with diverging first moment ($q_{\textrm{true}}=1.6$). 
Figure~\ref{fig} shows, for the three cases, $\overline{\ell}_q(\psi^*_q)$ (blue line) and $-S_1[P_q(\psi^*_q)]$ (orange line) as functions of $q$. The black dot indicates the intersection between the two curves, which identifies the estimated value $q^*$ where Eq.~\eqref{1oss_q} is realized.
The true values of the parameters and their inferred ML estimates $(q^*,\psi^*_{q^*})$ are presented in Table \ref{tab:table1}.
Since the left plot corresponds to $q_\textrm{true}=1$, it is a standard exponential distribution.
In such a case, the two curves intersect only for $q=1$. 
By contrast, the other two cases correspond to $q_\textrm{true}\ne 1$ and the two curves intersect in two points, namely $q=1$ and $q=q_\textrm{true}$.
In these cases, both intersections are solutions of Eq.~\eqref{1oss_q}, but the solution $q\ne 1$ is the one that corresponds to higher log-likelihood (and lower entropy).
This example is very simple but explanatory: it shows directly how Shannon entropy plays a role in model selection even when the distribution taken into consideration comes from the GMEP and maximizes R\'enyi, not Shannon.
We also stress once more that, in the last case, constraining the usual mean rather than the $q$-mean would have not been appropriate, since for $q > 1.5$ the usual mean diverges as $M\to\infty$; instead, by using the $q$-average, it becomes  possible to consistently characterize the original infinite-mean power-law distribution.\\

%%%%%%%%%%%%%
\begin{table}[b]
\caption{\label{tab:table1}%
Comparison of true parameters' values with ML estimates.
}
\begin{ruledtabular}
\begin{tabular}{lcdr}
\textrm{$q_{\textrm{true}}$}&
\textrm{$\psi_{\textrm{true}}$}&
\textrm{$q^*$}&
\textrm{$\psi^*_{q^*}$}\\
\colrule
1.0 & 5.0 & 1.0 & 5.0\\
1.3 & 3.0 & 1.3 & 2.9\\
1.6 & 7.0 & 1.6 & 7.3\\
\end{tabular}
\end{ruledtabular}
\end{table}
%%%%%%%%%%%%%%%

Our second and last example is the simple case of a system characterized by a Bernoulli random variable $C(G)$ taking value $C(G)=1$ with true underlying probability $p_\textrm{true}$, and value $C(G)=0$ with probability $1-p_\textrm{true}$. Constraining the $q$-average yields \begin{equation}
    p_q(G_i,\psi) = \frac{[1-(1-q)\,\psi\cdot C(G_i)]_+^{1/(1-q)}}{1+[1- (1-q)\,\psi ]^{1/(1-q)}}.
\end{equation}
Let us now call $p_q(\psi)$ the probability $p_q(G,\psi)$ when $C(G)=1$ and $1-p_q(\psi)$ the probability $p_q(G,\psi)$ when $C(G)=0$. 
It is easily verified that 
\begin{equation}
\langle C \rangle = p_q(\psi)
\end{equation}
and 
\begin{equation}
\langle C \rangle_q=\frac{p^q_q(\psi)}{p^q_q(\psi)+[1-p_q(\psi)]^q}.
\end{equation}
If we now consider $M$ i.i.d. realizations $\{C^*_m\}_{m=1}^M$ of $C$  and apply Eq.~\eqref{MLCq}, we get $p^{q^*}_{q^*}(\psi^*_{q^*})= p^{q^*-1}_{q^*}(\psi^*_{q^*})f_1$ where $f_1=\sum_{m=1}^M C^*_m/M$ is the empirical frequency of the observed instances where $C^*_m=1$. 
This relation trivially reduces to 
\begin{equation}
    f_1=p_{q^*}(\psi^*_{q^*}).
\end{equation}
Since there are infinite couples of $(\psi^*_{q^*},q^*)$ that satisfy the ML condition and produce exactly the same maximized log-likelihood, none of them has to be preferred over the other. 
According to our approach, one finds a result which recalls the Shannonian case: for a Bernoulli random variable, the parameters of the maximum entropy distribution have to be set so that the estimated probability matches the empirical frequency. 
This can be done for any value of $q$ and is therefore a degenerate case where no specific value of $q$ can be learned from the data, because the resulting maximum-entropy distributions are all identical to each other. This is not unexpected: in fact, what we have done here in practice is trying to capture the properties of a one-parameter binary random variable with a distribution that depends on two parameters.

\section{Conclusions\label{conclusions}}

A large body of literature has discussed the generalized axiomatic definition of entropy deriving from the relaxation (or unrestricted interpretation) of some of the \emph{SK} and \emph{SJ} axioms (in particular, \emph{SK4} and \emph{SJ3}). 
It is known that, when generalized in that way, the definition of entropy leads to parametric entropy families where a specific value of the entropic parameter(s) usually retrieves the ordinary Shannon functional.
In a maximum-entropy approach, each entropy family leads to a corresponding family of maximum-entropy probability distributions, indexed again by the entropic parameter(s), that provide the least biased inference about a system for which only limited information is available, in the form of empirical observations of a quantity treated as a soft constraint.
Unfortunately, when the estimated maximum-entropy distribution is `put back’ into its defining generalized entropy, a number of inconsistencies typically arise, including incompatibility with the ML principle, impossibility of determining the value of the entropic parameter(s) purely from empirical data, and disconnection from Shannon entropy when multiple independent observations of the same system are available.\\

In this paper, based on the fact that every member of an entropy family is ultimately intended as a quantification of the uncertainty encoded in the input probability distribution, we have introduced an uninformativeness axiom demanding that the maximally uncertain (i.e. uniform) probability distribution should always return the same (maximal) value of the entropy, irrespective of the value of the entropic parameters.
This simple axiom implies that all entropies take values within the same interval $[0,\ln\Omega]$, where $\Omega$ is the number of possible (unconstrained) microstates of the system, thereby equipping generalized entropies with a universal scale and meaning. 
The axiom considerably restricts the admissible members of entropy families. In particular, for both the \emph{UJK} and \emph{HT} entropies, the axiom selects only R\'enyi entropy as viable. A notable counterexample, dismissed by the axiom, is Tsallis entropy.
From an inferential point of view, the axiom guarantees that completely uninformative data (or equivalently the complete absence of empirical information) cannot be used to learn the value of entropic parameters.
At the same time we have showed that, when informative data are available, a straightforward extension of the ML principle leads to the optimal estimation of the entropic parameter(s), purely from empirical observations and without making any assumptions.\\

The resulting generalized ML approach couples the determination of the entropic parameters with that of the other structural parameters (Lagrange multipliers) of the maximum-entropy distribution. 
In particular, while the ML condition for the Lagrange multipliers indicates which specific combination of $M$ independent observations should be put equal to the generalized mean value of the constraint, the one for the entropic parameters coincides with the requirement that the log-likelihood of the data equals minus Shannon entropy.
This remarkable result shows that the connection between Shannon entropy and log-likelihood holds true also for generalized entropies (for the appropriate ML value of the entropic parameter) and is consistent with the assumed independence of the $M$ observations.
When $M=1$, the maximum-entropy probability returns coinciding values of R\'enyi and Shannon entropies, even if it maximizes the former but not the latter. 
For multiple independent observations ($M>1$), the connection between log-likelihood and Shannon entropy remains, while the connection with R\'enyi entropy disappears, as a result of independence. Therefore the log-likelihood, when maximized also over the entropic parameters, automatically finds the correct entropy to be used for model fitting and selection.\\

We believe the introduction of the uninformative axiom has beneficial effects for statistical inference and its many applications, and offers a way of constructing generalized entropies that have still controllable and consistent properties.

\bibliography{biblio}% Produces the bibliography via BibTeX.

\end{document}